# Research on fault diagnosis of nuclear power one-two circuit based on hierarchical multi-granularity classification network

CHEN Jiangwen，LI Siwei， GUO Jiang，CHENG Dongzhen，LIN Hua，WANG Wei

**Abstract:** The safe and reliable operation of complex electromechanical systems in nuclear power plants is crucial for the safe production of nuclear power plants and their nuclear power unit. Therefore, accurate and timely fault diagnosis of nuclear power systems is of great significance for ensuring the safe and reliable operation of nuclear power plants. The existing fault diagnosis methods mainly target a single device or subsystem, making it difficult to analyze the inherent connections and mutual effects between different types of faults at the entire unit level. This article uses the AP1000 full-scale simulator to simulate the important mechanical component failures of some key systems in the primary and secondary circuits of nuclear power units, and constructs a fault dataset. Meanwhile, a hierarchical multi granularity classification fault diagnosis model based on the EfficientNet large model is proposed, aiming to achieve hierarchical classification of nuclear power faults. The results indicate that the proposed fault diagnosis model can effectively classify faults in different circuits and system components of nuclear power units into hierarchical categories. However, the fault dataset in this study was obtained from a simulator, which may introduce additional information due to parameter redundancy, thereby affecting the diagnostic performance of the model.

The safe and reliable operation of complex electromechanical systems of nuclear power units is a necessary condition for the safety production of nuclear power plants and nuclear power plants. Due to the large number of mechanical and electrical equipment, sensing and monitoring points and related various instrument and control system parameters of the nuclear power unit plant and station, it is difficult to locate the unit in time when the fault occurs, resulting in large economic losses [1]. The existing fault diagnosis methods mainly focus on a single equipment or subsystem, and it is difficult to analyze the internal relationship and mutual influence between different types of faults at the whole unit level. Therefore, it is of great significance to carry out fine hierarchical classification diagnosis of nuclear power unit faults. EfficientNet applies an efficient convolutional neural network architecture to achieve efficient model design by uniformly scaling the depth, width and resolution of the network, thereby reducing the number of calculations and parameters while maintaining accuracy. It is a solution to obtain high performance in the case of limited computing resources. Specifically designed for image classification and recognition [2]. Based on the fault data set of the full scope simulator of nuclear power plant, the EfficientNet large model is applied to the hierarchical multi-granularity classification diagnosis of nuclear power unit faults, and compared with several commonly used large models, so as to explore the application prospect of EfficientNet in the field of nuclear power plant fault diagnosis.

## 1　EfficientNet

Large models require a large amount of computing resources and large-scale data for training. For small and medium-sized institutes or institutions, the cost of developing or training their own large models from scratch is difficult to accept. This makes it difficult to deploy large models in specific areas of industry. Therefore, this study chooses to deploy the pre-trained large model EfficientNet and other models by transfer learning, which can also provide a feasible reference for other researchers to use large models.

Figure 1 shows the schematic diagram of the network structure of EfficientNet, where the detailed structure of Module1 to Module3 is shown in Figure 2, and the total number of network layers is 237. Among them, the convolution kernel size, resolution, number of channels and number of layers are shown in Table 1. EfficientNet's optimization goal is to maximize model accuracy given a resource budget, and the model proposes a new parameter scaling method, That is, a compound factor "ϕ" is used to uniformly scale the network width, depth, and resolution. The specific principle is as follows:

$$\text{depth: } d = \alpha^\phi$$
$$\text{width: } \omega = \beta^\phi$$
$$\text{resolution: } r = \gamma^\phi$$
$$\text{s.t. } \alpha \cdot \beta^2 \cdot \gamma^2 \approx 2; \alpha \geq 1, \beta \geq 1, \gamma \geq 1$$

Where α, β, γ are constants obtained from a small range of network searches, which determine how computing resources are allocated, and "ϕ" can be used to control the number of resources used. Since regular convolutions are computationally proportional to d, $\omega^2$, $r^2$, doubling the depth doubles the FLOPS, and doubling the width and resolution increases the FLOPS by a factor of 4. In this paper, the arbitrary "ϕ added constraints $\alpha \cdot \beta^2 \cdot \gamma^2 \approx 2$, make the model as a whole increased the amount of calculation about 2 times ϕ.

This method works by finding the best coefficients for width, depth, and resolution, and then combining them together to fit the original network model, adjusting for each dimension. Scale the model from a holistic perspective. This compound scaling method can consistently improve the accuracy and efficiency of the model compared to the traditional method. Test results on existing models MobileNet and ResNet show that it improves the accuracy by 1.4% and 0.7%, respectively.

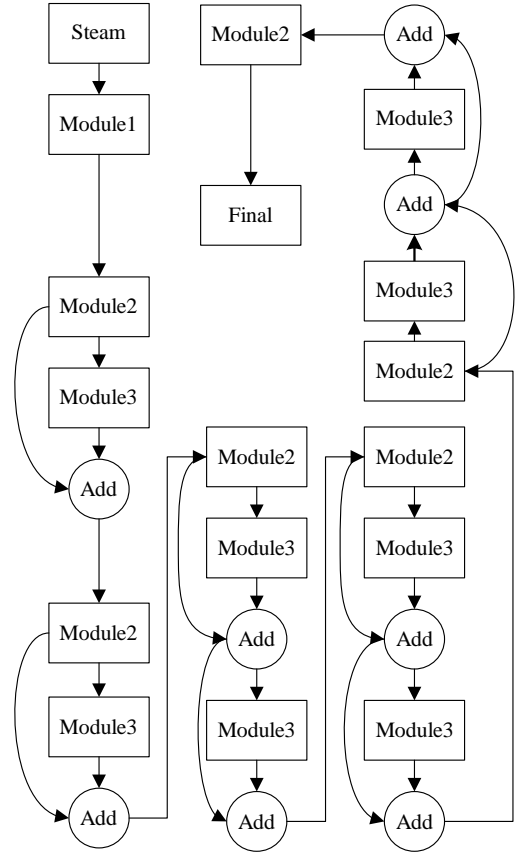

Fig.1　EfficientNet architecture principles

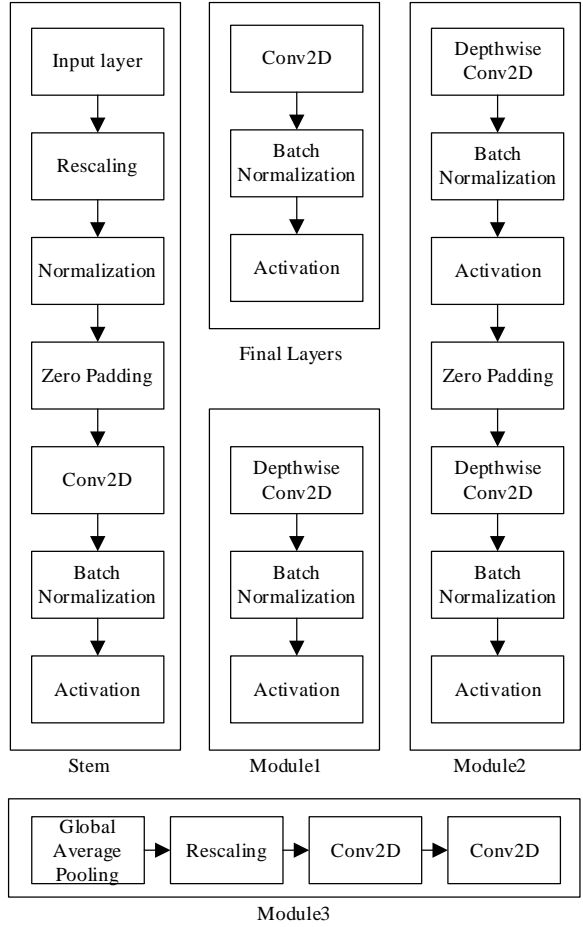

Fig.2 EfficientNet module structure

Table 1 Network parameter

| 阶段 i | 卷积核大小 $\widehat{F}_i$ | 分辨率 $\widehat{H}_i \times \widehat{W}_i$ | 通道数 $\widehat{C}_i$ | 层数 $\widehat{L}_i$ |
|---|---|---|---|---|
| 1 | Conv3x3 | 224×224 | 32 | 1 |
| 2 | MBConv1, k3x3 | 112×112 | 16 | 1 |
| 3 | MBConv6, k3x3 | 112×112 | 24 | 2 |
| 4 | MBConv6, k5x5 | 56×56 | 40 | 2 |
| 5 | MBConv6, k3x3 | 28×28 | 80 | 3 |
| 6 | MBConv6, k5x5 | 14×14 | 112 | 3 |
| 7 | MBConv6, k5x5 | 14×14 | 192 | 4 |
| 8 | MBConv6, k3x3 | 7×7 | 320 | 1 |
| 9 | Conv1x1 & Pooling & FC | 7×7 | 1280 | 1 |

## 2 Fault data set of the first and second circuits of nuclear power plants

At present, the research on fault diagnosis of nuclear power unit mainly focuses on a single equipment or subsystem of nuclear power unit, and the fault data only comes from the specific object under study. The deficiency of this kind of research is that it fails to comprehensively judge the fault type by comparing the overall system behavior changes under different fault states from the perspective of the whole unit operation state. To address this limitation, this study proposes a fault data construction method based on a full-scale nuclear power plant simulator to build a data warehouse for the fault diagnosis model.

The primary and secondary circuits of nuclear power units contain multiple subsystems. In this study, the reactor coolant system, main steam system, condensate system and main feed water system in the primary and secondary circuits of nuclear power units are taken as the object, and a total of 16 types of typical faults are diagnosed and studied.

By adding different fault conditions one by one in the nuclear power plant full scope simulator, it can be generated in real time in the unit DCS system

The total number of operating parameters to the unit at that time was 10725. These parameters contain multi-dimensional information such as thermal, hydraulic, electrical and logical control during unit operation, which can fully characterize the overall operation state of the unit when the fault occurs. The collected one-dimensional time series fault data is processed into fault pictures by image gray algorithm, which can better show the characteristic differences between different faults.

The commonly used methods for two-dimensional data include wavelet transform (WT), Gramian Angle Field method (GAFS), Markov transition Field method (MTF), image gray change method, etc. [3]. These methods transform the one-dimensional time of a single parameter into two dimensions in the time dimension, and can obtain the periodicity and autocorrelation characteristics of a single parameter over time, but cannot obtain the overall representation and overall characteristics among nuclear power plant parameters [4].

The data (10725 parameter data in the DCS system) were processed into a two-dimensional matrix using the image gray algorithm. The size of the matrix is 104*104, which is used to store all parameter data. The unfilled positions in the matrix are supplemented with 0. Then the matrix data is normalized to the interval [0,255], and finally it is saved and output as a grayscale image. FIG. 3 shows the grayscale images of some typical faults in the critical system of the first and second loops.

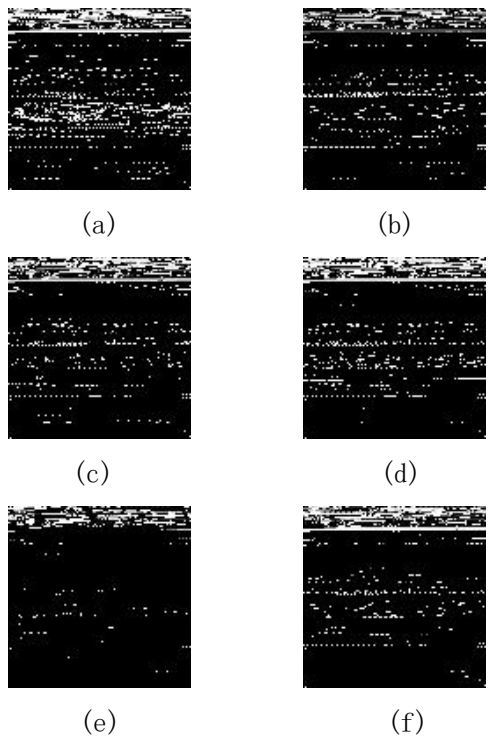

Fig.3　6 typical fault pictures

## 3　基于 EfficientNet 的分层多粒度分类模型

Figure 4 shows the structural principle of the hierarchical multi-granularity classification model based on EfficientNet.

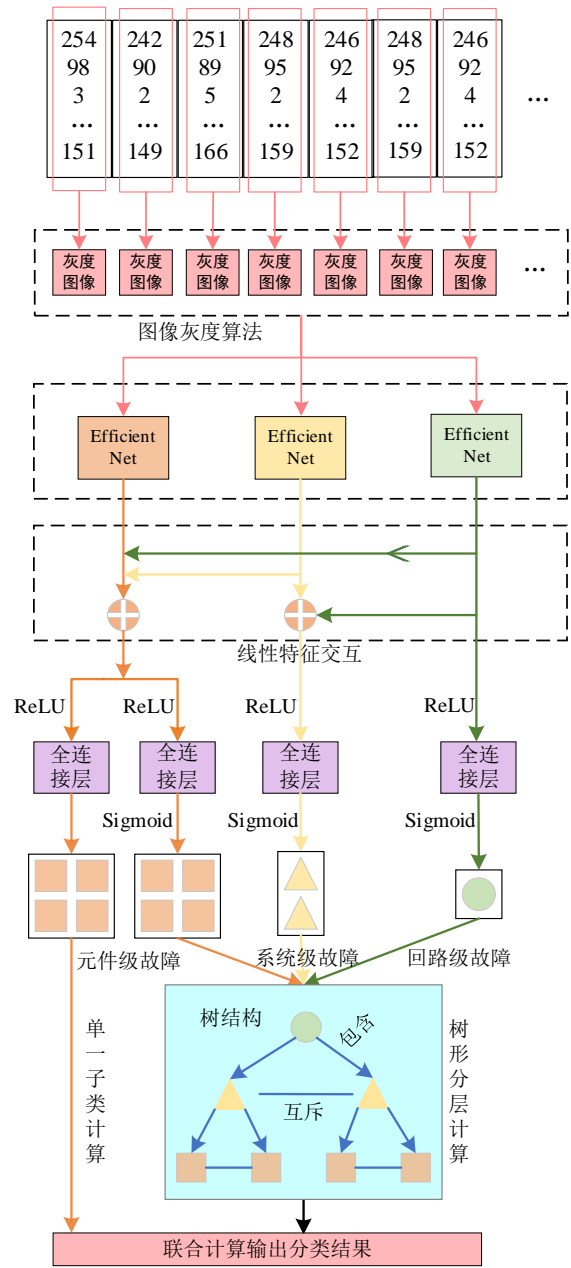

Fig.4　Fault diagnosis model

After the gray image of the fault of the nuclear power unit is diagnosed by the 3-level EfficientNet large model, the initial judgment results of three levels of loop, system and component are given. Then, the features of faults at different levels are shared through linear combination, and the nonlinear transformation ReLU is applied to the combined features [5]. Therefore, nuclear electronic-grade component faults not only have unique properties, but also inherit properties of system and loop faults of their parent class. Finally, the diagnosis probability of fault samples was given by jointly analyzing the results of single classification and tree hierarchical calculation, and the fault diagnosis process was completed. The 16 types of faults in Table 2 are sampled under three operating conditions of the nuclear power unit: 10% start-up power operation, 50% and 100% power steady state operation. The sampling interval was 1 second and the sampling period was 20 minutes. The total fault data set consists of 57,600 fault images. The constructed dataset was divided into training set, test set and validation set according to the ratio of 8:1:1. Figure 5 shows the training accuracy curves of the EfficientNet-based hierarchical multi-granularity classification fault diagnosis model with VGG16 and ResNet50.

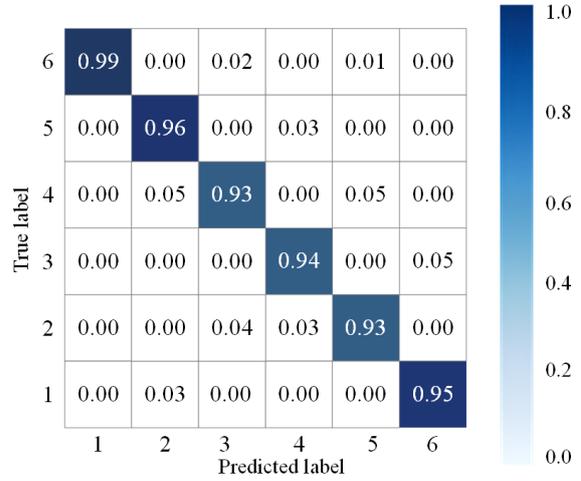

Fig.6   Confusion matrix classification results

It can be seen that for the frequency conversion failure of the main pump MP01A, the accuracy of the model reaches 99%, and the average accuracy exceeds 95%, which proves that the model can effectively realize the hierarchical classification of faults.

## 4 Model diagnosis example verification

To better evaluate the performance of the fault diagnosis model in this study, it is compared with other hierarchical multi-granularity classification methods, including HMCN proposed by Wehrmann et al. [6] and C-HMCNN proposed by Giunchiglia et al. [7]. The results are recorded in Table 2。

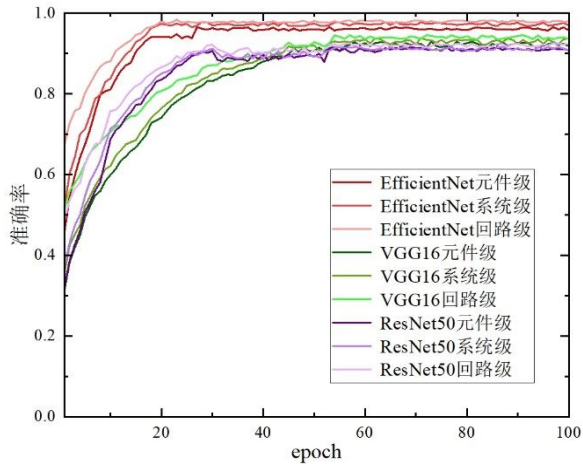

Fig.5   Different model training results

It can be seen that the hierarchical multi-granularity classification fault diagnosis model based on EfficientNet has an accuracy of 96.03 on the training set, which is 3.28% and 2.15% higher than the training accuracy of VGG16 and ResNet50, respectively.

Due to space limitation, this study selects 6 types of primary and secondary loop typical faults out of 16 types of faults, and gives the confusion matrix of their classification, as shown in FIG. 6. Labels 1 to 6 in the figure are, respectively, large break in CL1 cold pipe section, misstart of safety injection system, break in SG1 steam pipe, break in SG1 feed pipe, drop of control rod H08 and frequency conversion failure of main pump MP01A.

Table 2   Model results

| Loop | Hierarchy | Wehr-mann's | Giunc-higlia's | Ours |
|---|---|---|---|---|
| Loop1 | Root | 95.35% | 95.24% | 97.22% |
|  | Parent level | 91.87% | 94.54% | 96.85% |
|  | Child level | 80.41% | 82.58% | 96.23% |
| Loop2 | Root | 96.72% | 97.34% | 97.85% |
|  | Parent level | 96.43% | 96.15% | 97.13% |

| | | | |
|---|---|---|---|
| Child level | 94.24% | 94.59% | 94.96% |

The test results on the validation set show that the fault diagnosis performance of the fault diagnosis model proposed in this study is better at different levels of the first and second loops. The fault diagnosis system of nuclear power plant is established based on NuSim simulation platform, and the system interface is shown in Figure 7.

After the fault data of the nuclear power plant full-scale simulator is connected to the system, the system automatically performs the tasks of data analysis and fault diagnosis. As can be seen from FIG. 7, after the large break fault of CL1 cold pipe section is added to the simulator, the system can quickly identify the loop subsystem to which the fault belongs, and the confidence is more than 96%, which means that the diagnosis ability is good.

## 5 Conclusion

The hierarchical multi-granularity classification fault diagnosis model based on EfficientNet effectively realizes the hierarchical classification of nuclear power fault by using large model technology and hierarchical classification network, which provides new ideas for nuclear power fault diagnosis. This study collects sufficient fault data based on the full scope simulator of nuclear power plant, which effectively solves the problems of high cost of nuclear power fault data set construction, scarce number of faults and poor data quality. The EfficientNet large model is better than the traditional algorithm model in fault classification performance. At the same time, combined with the hierarchical classification network, the fault diagnosis model can better identify the fault level and granularity, which can improve the decision-making efficiency for nuclear power plant operators and reduce the risk of false diagnosis. However, due to the large amount of calculation, the balance between accuracy and real-time performance should be considered when the large model is applied in the field of nuclear power fault diagnosis.

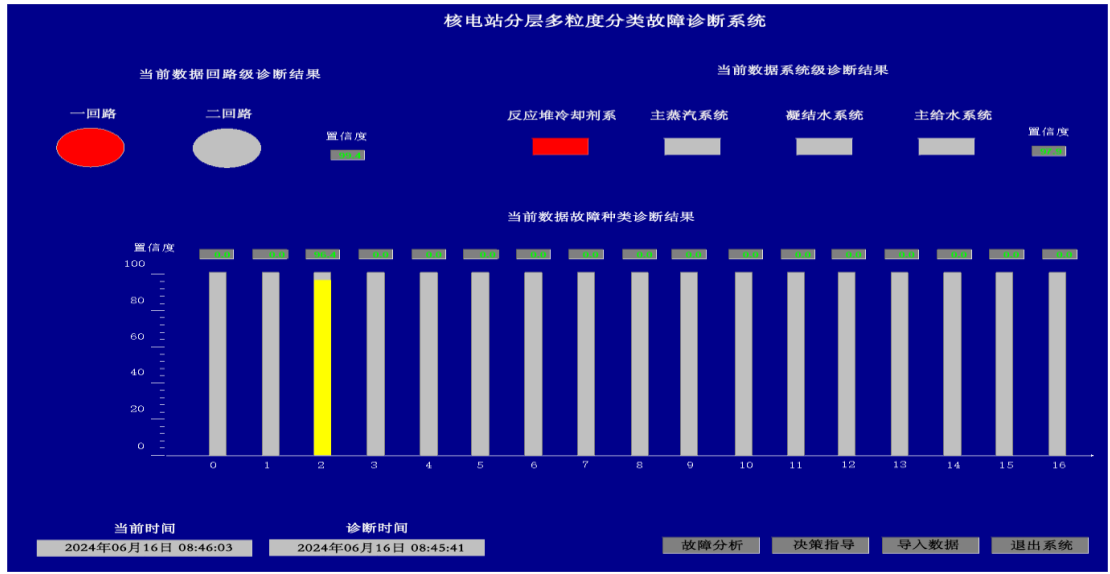

Fig.7 Fault diagnosis result of large break in CL1 cold pipe section